\documentclass[preprint,aps,showpacs,nofootinbib]{revtex4}

\usepackage{graphicx}
\usepackage{epsfig,amssymb,amsmath,color}
\usepackage{epstopdf}

\newcommand{\beq}{\begin{equation}}
\newcommand{\eeq}{\end{equation}}
\newcommand{\bea}{\begin{eqnarray}}
\newcommand{\eea}{\end{eqnarray}}
\newcommand{\bear}{\begin{array}}
\newcommand {\eear}{\end{array}}
\newcommand{\bef}{\begin{figure}}
\newcommand {\eef}{\end{figure}}
\newcommand{\bec}{\begin{center}}
\newcommand {\eec}{\end{center}}
\newcommand{\non}{\nonumber}

\newcommand{\la}{\left\langle}
\newcommand{\ra}{\right\rangle}

\def\EQ#1{Eq.~(\ref{#1})}

\def\GEV#1{10^{#1}{\rm\,GeV}}

\def\lrf#1#2{ \left(\frac{#1}{#2}\right)}
\def\lrfp#1#2#3{ \left(\frac{#1}{#2} \right)^{#3}}

\begin{document}
\draft
\tighten
\preprint{KEK-TH-1710
}
\title{\large \bf
Multi-Natural Inflation in Supergravity
}
\author{
Michael Czerny\,$^{a,\ast}$\footnote[0]{$^\ast$ email: mczerny@tuhep.phys.tohoku.ac.jp},
Tetsutaro Higaki\,$^{b,\star}$\footnote[0]{$^\star$ email: thigaki@post.kek.jp},    
Fuminobu Takahashi\,$^{a,\,c\,\dagger} $\footnote[0]{$^\dagger$ email: fumi@tuhep.phys.tohoku.ac.jp}
    }
\affiliation{
    $^a$ Department of Physics, Tohoku University, Sendai 980-8578, Japan \\
    $^b$ Theory Center, KEK, 1-1 Oho, Tsukuba, Ibaraki 305-0801, Japan \\
    $^c$ Kavli IPMU, TODIAS, University of Tokyo, Kashiwa 277-8583, Japan
    }

\vspace{2cm}

\begin{abstract}
We show that the recently proposed multi-natural inflation can be realized within the 
framework of 4D ${\cal N}=1$ supergravity. The inflaton potential mainly consists of two  sinusoidal potentials 
that are comparable in size, but have different periodicity with a possible non-zero relative phase. 
For a sub-Planckian decay constant, the multi-natural inflation model is reduced to axion hilltop inflation.
We show that, taking into account the effect of the relative phase, the spectral index can be increased
to give a better  fit to the Planck results, with respect to the hilltop quartic inflation.
We also consider a possible UV completion based on a string-inspired model. Interestingly, the Hubble parameter
during inflation is necessarily smaller than the gravitino mass, avoiding possible moduli destabilization.
Reheating processes as well as non-thermal leptogenesis are also discussed. 
\end{abstract}
\pacs{}
\maketitle


\section{Introduction}

The recent observations of the cosmic microwave background (CMB) by the Planck satellite \cite{Ade:2013uln} 
showed that $\Lambda$CDM cosmology is consistent with the data
and fluctuations in the cosmic microwave background (CMB) can be explained by single-field inflation
\cite{Guth:1980zm,Linde:1981mu},
which solves the fine-tuning problems in the early universe.
The spectral index $n_s$ and the tensor-to-scalar ratio $r$ are tightly constrained by
the Planck data combined with other CMB observations \cite{Ade:2013uln}:
\bea
\label{planck}
n_s &=& 0.9603 \pm 0.0073, \\
r &<& 0.11 ~(95\%~{\rm CL}).
\eea
The index $n_s$ is determined by the shape of the inflaton potential, whereas the ratio $r$ is done by
the energy scale of the potential:\footnote{After submission of this paper, the BICEP2 collaboration announced the detection
of the primordial B-mode polarization, which can be explained by $r = 0.20^{+0.07}_{-0.05}$~\cite{Ade:2014xna}. See note added at the end of this paper.}
\bea
H_{\rm inf} \simeq 8.5 \times 10^{13}~{\rm GeV} 
\left( \frac{r}{0.11} \right)^{1/2}.
\label{stratio}
\eea
Here, $H_{\rm inf}$ is the Hubble scale during inflation.

To construct a viable inflation model, the inflaton potential should be under good control so as not to break
the slow-roll condition and to suppress the inflation scale compared with the the Planck scale.
There have been many attempts to accomplish this. 
One way is to introduce a certain symmetry which keeps
the inflaton potential flat. See Refs.~\cite{Linde:1983gd,Kawasaki:2000yn,Kawasaki:2000ws,
Kallosh:2010ug,Kallosh:2010xz,Takahashi:2010ky,Nakayama:2010kt,Harigaya:2012pg,Silverstein:2008sg,McAllister:2008hb,Kaloper:2008fb,Croon:2013ana,
Nakayama:2013jka,Ellis:2013xoa,Kallosh:2013pby}  for various chaotic inflation models along this line.
In this sense, an axion is a good candidate for the inflaton due to the approximate shift symmetry 
\bea
\phi \to \phi + {\rm const}. ,
\eea
which controls its potential structure and suppresses the scale of inflation to be consistent with observation.
Here, $\phi$ is an axion. So, it is possible to consider a natural inflation model \cite{Freese:1990ni,Adams:1993ni}
with an axion potential $V(\phi)$ given by:
\bea
V(\phi) = \Lambda^4 \bigg[1 - \cos\left(\frac{\phi}{f}\right) \bigg] .
\eea
In this model, the shift symmetry is broken non-perturbatively 
by the dynamical scale $\Lambda$ much smaller than the Planck scale.
However, the decay constant $f$ is required to be larger than the Planck scale\footnote{
See \cite{Kim:2004rp} for realizing a large decay constant effectively, 
\cite{Mohanty:2008ab,Germani:2010hd} for other ways to relax the bound on the decay constant,
and \cite{Kaplan:2003aj,BlancoPillado:2004ns,Dimopoulos:2005ac,Silverstein:2008sg,McAllister:2008hb} 
for other models with axion(s).
}, 
$f \gtrsim 5M_{\rm Pl}$ \cite{Ade:2013uln}, for the predicted $n_s$ and $r$ to be consistent with the observed values,
where $M_{\rm Pl} \simeq 2.4 \times 10^{18}$GeV is the reduced Planck mass.\footnote{
Hereafter, we take the Planck unit of $M_{\rm Pl}=1$ for a simplicity, unless otherwise stated.
}
So, one might worry about the control of the correction $f/M_{\rm Pl}$ after all.

Recently, two of the present authors (MC and FT) proposed an extension of natural inflation,
called multi-natural inflation~\cite{Czerny:2014wza}, in which the inflaton potential mainly consists of 
two (or more) sinusoidal functions. Interestingly, multi-natural inflation is versatile enough to 
realize both large-field and small-field inflation. In the case of large-field inflation with super-Planckian 
decay constants, the predicted values of the spectral index as well as 
the tensor-to-scalar ratio can be closer to the center values of the Planck results, with respect 
to the original natural inflation. 

In the case of small-field inflation with sub-Planckian decay constants,
we arrange those sinusoidal functions so that they conspire to make the inflaton potential sufficiently flat
for slow-roll inflation. 
In a certain limit, this axion hilltop inflation is equivalent to hilltop quartic inflation \cite{Linde:1981mu}.
The hilltop quartic inflation has been studied extensively so far, and it is known that, for the e-folding number $N_e \simeq 50$, 
 its predicted spectral index tends to be too low to explain  the Planck results [\ref{planck}]. 
There are various proposals for resolving this tension in the literature.

The purpose of this paper is twofold.
First, we will show that the predicted spectral index for the axion hilltop inflation can be increased
with respect to the hilltop quartic inflation case by including a relative phase between 
two sinusoidal functions. This gives a better fit to the Planck data.
Second, we consider a UV completion of multi-natural inflation within supergravity(SUGRA)/string theory.
This is because a viable inflation model can be easily realized for large decay constants close to the GUT or 
Planck scale, and because the string theory offers many axions~through compactifications
\cite{Svrcek:2006yi,Blumenhagen:2006ci,Arvanitaki:2009fg}, 
some of which could play an important role in inflation. Also, non-perturbative dynamics which explicitly break
the axionic shift symmetry can be studied rigorously in a supersymmetric (SUSY) framework.

The rest of this paper is organized as follows. In Sec.\ref{SecSUGRA} we build an axion hilltop inflation model
with one axion multiplet in the context of SUGRA, taking into account SUSY breaking effects.
In Sec.\ref{Secstring}, we consider a UV completion of multi-natural inflation in the string-inspired model,
which is reduced to the model analyzed in Sec.\ref{SecSUGRA} in the low energy effective theory.
The last section is devoted to discussion and conclusions. 

\section{Multi-natural inflation in SUGRA}
\label{SecSUGRA}
\subsection{Setup}
Let us consider a supergravity realization of multi-natural inflation in which an axion plays the role
of the inflaton~\cite{Czerny:2014wza}. 
To this end, we introduce the following K\"ahler and super-potentials with an axion chiral superfield $\Phi$,
\bea
K &=& K(\Phi + \Phi^{\dag}), \\
W &=& W_0 + Ae^{-a\Phi} + Be^{-b\Phi},
\label{W1}
\eea
where  $a>0$, $b>0$ and $a \ne  b$.\footnote{
The case of the irrational ratio of $a/b$ was considered in Ref.~\cite{Banks:1991mb}.
The following discussion holds even in this case. 
} In the following we assume $|a^4A| < |b^4B|$ without loss of generality. 
The scalar potential is given by
\bea
\label{sugra}
V= e^{K}[K^{i\bar{j}}(D_iW) (\overline{D_jW})-3|W|^2],
\eea
with $D_i W = (\partial_i K) W + \partial_i W$.
For convenience we write the scalar component of $\Phi$ as
\bea
\Phi = \sigma + i \varphi,
\eea
where $\sigma$ and $\varphi$ are the saxion and the axion, respectively. 
Note that the K\"ahler potential respects the axionic shift symmetry,
\bea
\varphi \to \varphi+ {\rm const},
\eea  
which is explicitly broken by the two exponentials in the superpotential. 
We assume that the breaking of the shift symmetry is so weak that the axion mass is
hierarchically smaller than the saxion mass. As we shall see shortly,
this is the case if
\bea
|A|,  |B| \ll  |W_0| < 1.
\label{para1}
\eea
The saxion is then decoupled from the inflaton dynamics. 
The softly broken shift symmetry
is one of the essential ingredients for multi-natural inflation, and such a small breaking naturally arises
from non-perturbative effects at low-energy scales\footnote{Such softly broken shift symmetries and the associated light axions 
may play an important cosmological role in a different context. For instance, the axion 
with mass $7$\,keV~\cite{Higaki:2014zua} can explain the recently
observed X-ray line at about $3.5$\,keV~\cite{Bulbul:2014sua,Boyarsky:2014jta}.}.
On the other hand, if the saxion mass were comparable to the axion, one would have to follow the multi-field dynamics during inflation.
Although  the analysis becomes rather involved, it is possible to realize successful inflation as is the case with 
racetrack inflation~\cite{BlancoPillado:2004ns}.

\subsection{Saxion stabilization}
First we study the saxion stabilization in the above setup.
For simplicity and concreteness we consider the following K\"ahler potential  to stabilize the saxion:
\bea
K =  \frac{f^2}{2}(\Phi+\Phi^\dag)^2
\label{K2}
\eea
with $f \lesssim 1$.
We will see in the next section that the K\"ahler potential of this form is indeed obtained in the 
low energy effective theory of a more realistic string-inspired model.

The kinetic term for the saxion and the axion is given by
\beq
{\cal L}_{\rm kin} \;=\; K_{\Phi {\bar \Phi}} \partial \Phi^\dag \partial \Phi =  f^2(\partial \sigma)^2 + f^2(\partial \varphi)^2.
\eeq
For the moment let us focus on the saxion stabilization by setting $A=B=0$. Then the saxion potential has
a $Z_2$ symmetry, $\sigma \rightarrow -\sigma$, and therefore the saxion potential has an extremum at
the origin $\sigma=0$. In fact, the origin can be the potential minimum as shown below. 

The saxion potential is approximately given by
\bea
V &= & e^{2f^2 \sigma^2}\bigg(4 f^2 \sigma^2 - 3  \bigg) |W_0|^2 + \Delta V \\
& \simeq &   2 f^2  |W_0|^2 \sigma^2 + \cdots,
\eea
where we have expanded the potential around the origin in the second equality, and we have added
 a sequestered uplifting potential $\Delta V$ to cancel the cosmological constant\footnote{
For instance, one can consider
$K= -3\log \left[e^{-K(\Phi + \Phi^{\dag})/3}-\{XX^{\dag} - (X X^\dag)^2/\Lambda^2 \}/3 \right]$ 
and $W= W_0 + \sqrt{3}W_0 X + W(\Phi)$ 
to break the SUSY and to obtain a small cosmological constant. In this model the vacuum and the mass are given by 
$\langle X \rangle \sim \Lambda^2 \ll 1$ and $m_{3/2}/\Lambda \gg m_{3/2}$ \cite{Kitano:2006wz}.
Then, $\Delta V = e^{K}|D_X W|^2K^{X \bar{X}} = 3|W_0|^2 e^{2K(\Phi + \Phi^{\dag})/3} $. 
The SUSY breaking fields can be integrated out during inflation as long as its mass is heavier than or comparable to
the gravitino mass. 
See \cite{Kachru:2003aw,Choi:2005ge,Lebedev:2006qq,Dudas:2006gr} for related topics.
},
\bea
\Delta V &=& 3 e^{2K/3} |W_0|^2  
\simeq \bigg(3 + 4f^2  \sigma^2 + \cdots\bigg)|W_0|^2.
\eea 
Thus, the saxion is stabilized at the origin with mass $m_\sigma \simeq \sqrt{2} |W_0|$.
Note that the saxion is stabilized by the SUSY-breaking effect through the equation
\bea
\partial_{\Phi}K = 0.
\eea
The saxion can be similarly stabilized for a more general K\"ahler potential;
see Refs.~\cite{Conlon:2006tq,Choi:2006za,Higaki:2011me} for detailed discussions on the saxion stabilization.
In general, the saxion mass is considered to be of order the gravitino mass.

\begin{figure}[t!]
\begin{center}
\includegraphics[width=9cm]{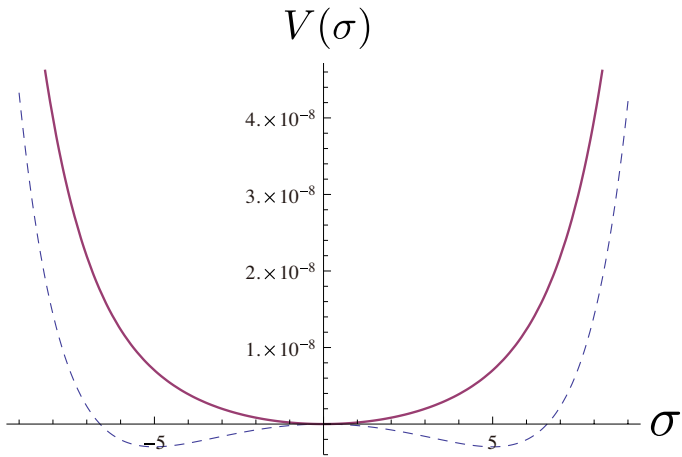}
\caption{The saxion potential for $A=2.3\times 10^{-12},~B=A/4,~a=2\pi/10,~b=2\pi/5,~f=0.1$ and $W_0=10^{-4}$. 
We have set $\varphi = 0$.  The dashed (blue) line shows the saxion potential  without the uplifting potential;
the saxion is stabilized at $\sigma \sim \pm 5$, where the vacuum energy $3|W_0|^2$ is added for visualization purpose.
The saxion can be stabilized near the origin if the sequestered uplifting potential $\Delta V$ is added, as shown by
the solid (red) line. 
}
\label{fig1}
\end{center}
\end{figure}

The axion mass is protected by a shift symmetry.  For a sufficiently small breaking of the shift symmetry, therefore, 
the axion acquires a mass much smaller than the saxion mass, while the saxion stabilization studied above remains 
almost intact. 
See Fig.\ref{fig1} for the saxion potential in the presence of small explicit breaking of the shift 
symmetry, with and without the uplifting potential.  We can see that the saxion is stabilized near the origin, when
the sequestered up-lifting potential is added. 
The saxion vacuum is located near the origin
as long as the parameters satisfy the relation
\bea
\frac{|A|}{|W_0|f^2} \sim \frac{|B|}{|W_0|f^2} \lesssim 10^{-2},
\eea
i.e., the axion mass is much lighter than that of the saxion by a factor of ten.

 We shall see that the Hubble parameter during inflation is necessarily smaller than the
gravitino mass as long as (\ref{para1}) is met. Then the saxion stabilization is hardly affected
by inflation, and we can integrate out the saxion during inflation. This makes the inflation dynamics 
extremely simple: the inflationary epoch is described by single-field inflation driven by the axion.

\subsection{Axion hilltop inflation}

Let us study the axion potential.
Using $U(1)_R$ symmetry and an appropriate shift of $\varphi$,  we can set $W_0$ and $A$ real and positive, while $B$ 
is complex in general. To take account of this complex phase, we replace $B$ with $B e^{- i \theta}$, where $B$ is a real and positive
constant, and $\theta$ represents the relative phase between the two exponentials. 
Using $\la \sigma \ra \simeq 0$, the axion potential can be approximately written as
\bea
\nonumber
V_{\rm axion}(\phi) &\simeq&
6 A W_0 \left[ 1 - \cos\left(\frac{\phi}{f_1}\right)\right] + 6 B W_0 \left[1- \cos\left(\frac{\phi}{f_2} + \theta \right)\right]  \\
&& - 2 A B \left( \frac{2}{f_1 f_2} -3\right) \left[
1- \cos\bigg[ \left( \frac{1}{f_1} -\frac{1}{f_2}  \right)\phi -\theta  \bigg] 
\right] + {\rm const},
\label{pot}
\eea
where $\phi \equiv \sqrt{2} f\varphi$ is the canonically normalized axion field, and we have defined
\bea
f_1 \equiv \frac{\sqrt{2}f}{a},~~~f_2 \equiv \frac{\sqrt{2}f}{b},
\eea
with $f_1 \ne f_2$.
We have added the vacuum energy from SUSY breaking to obtain the Minkowski spacetime in the true vacuum 
and the last constant term in \EQ{pot}  depends on $\theta$; it vanishes for $\theta =0$.

We impose relation (\ref{para1}) to realize a hierarchy between the saxion mass and the axion mass.
Then, the third term in \EQ{pot} becomes irrelevant\footnote{
In the presence of two axion fields, the third term can be responsible for natural inflation with an effective super-Planckian 
decay constant~\cite{Kim:2004rp}.
} and the first two terms are equivalent to the inflaton potential
for the multi-natural inflation discussed in Ref.~\cite{Czerny:2014wza}. A successful multi-natural inflation requires the two sinusoidal functions to have comparable magnitude and periodicity, i.e.,
\bea
&&A \sim B \ll W_0,\\
&&f_1 \sim f_2.
\eea
The inflation scale is roughly given by $H_{\rm inf} \sim (A W_0)^{1/2} \ll W_0$, and so, the saxion mass is generically heavier
than the Hubble parameter during inflation, which justifies our assumption. 
Also, the upper bound on the tensor mode (\ref{stratio}) places the following condition,
\bea
AW_0  \lesssim 10^{-10},
\eea
for $f_1 \sim f_2  \lesssim {\cal O}(1)$.

The inflaton potential must have a flat plateau in order to realize successful inflation with sub-Planckian
decay constants. In the following we show that, in a certain limit, the axion potential is reduced to the hilltop quartic inflation
model. For the moment we focus on the first two terms in  \EQ{pot}. In the numerical calculations we will
include all the terms. 

Requiring that the first, second and third derivatives of $V_{\rm axion}$ vanish  and
the fourth derivative of $V_{\rm axion}$ is negative at $\phi = \phi_{\rm max}$,
we obtain the following conditions among the parameters:
\begin{align}
 \label{cond1}
\sin\left(\frac{\phi_{\rm max}}{f_1}\right) &= \sin\left(\frac{\phi_{\rm max}}{f_2} + \theta \right)  = 0,\\
 \label{cond2}
-\cos\left(\frac{\phi_{\rm max}}{f_1}\right) &=  \cos\left(\frac{\phi_{\rm max}}{f_2} + \theta \right)  = 1,\\
 \label{cond3}
 \frac{A}{f_1^2} &= \frac{B}{f_2^2},
\end{align}
where we have used our assumption $a^4 A < b^4 B$, i.e., $A/f_1^4 < B/f_2^4$, to fix the sign of the cosine functions. 
Note that $A > B$ as well as $f_1 > f_2$ must be satisfied to meet the above conditions. 
We choose the following solutions without loss of generality, 
\bea
\label{sol1}
\phi_{\rm max} &=& \pi f_1,\\
\label{sol2}
\theta &=& - \pi \frac{f_1}{f_2}~~~~({\rm mod}~2\pi)
\eea
The inflaton potential becomes simple for a particular choice of $f_1 = 2f_2$ (i.e. $A=4B$), as the
relative phase $\theta$ vanishes. 

\begin{figure}[t!]
\begin{center}
\includegraphics[width=7.9cm]{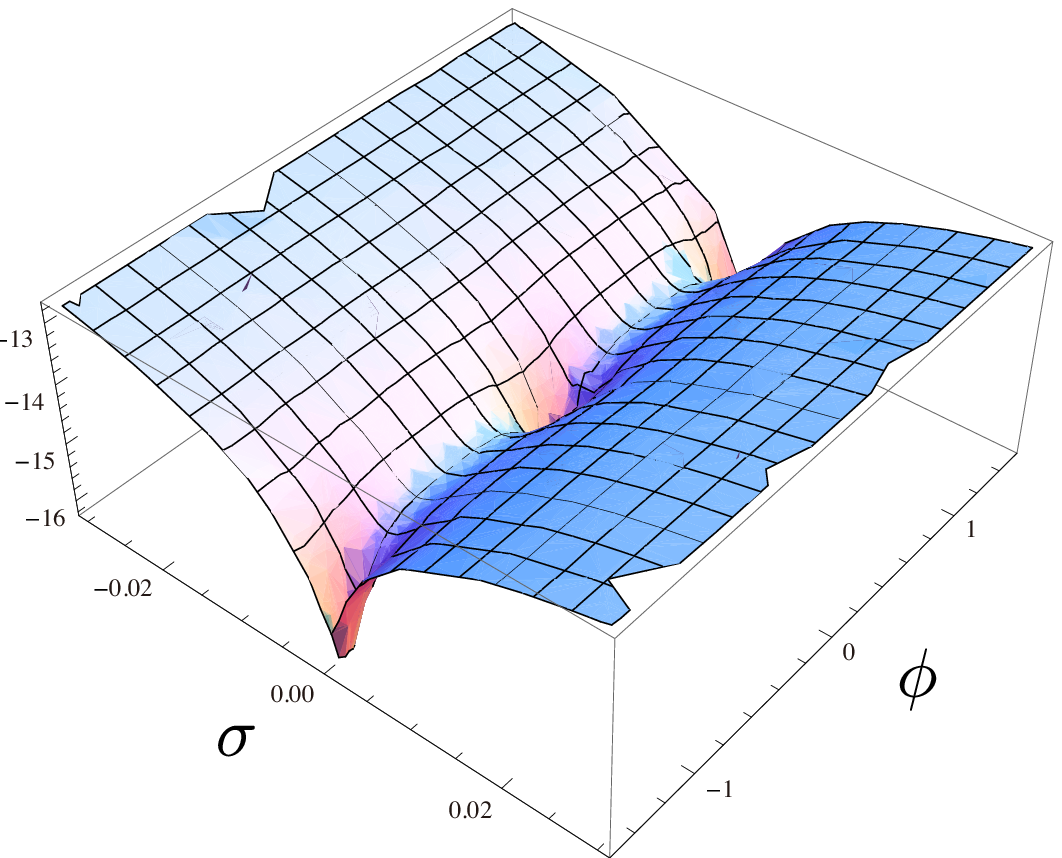}
\includegraphics[width=7.9cm]{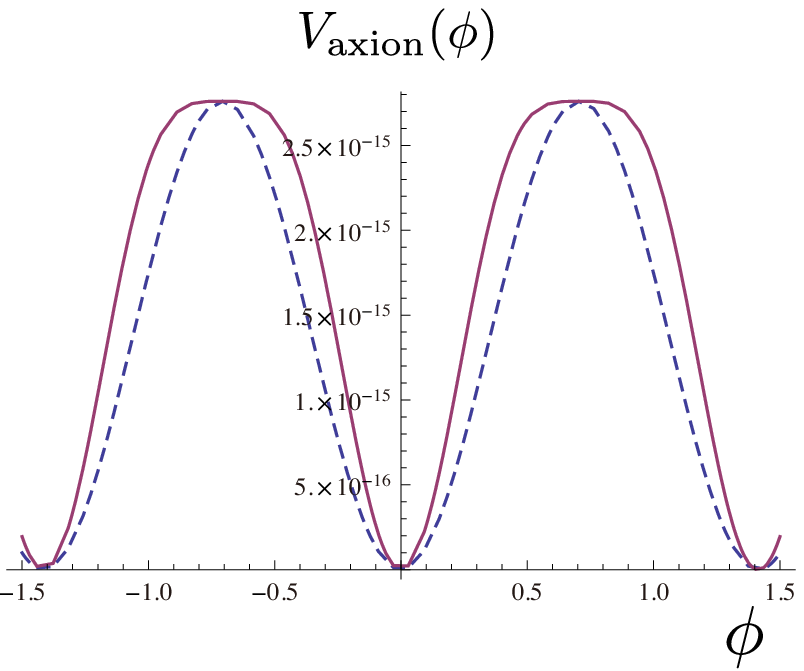}
\caption{
The scalar potential for the saxion and the axion (left) and the axion potential at the section of $\la \sigma \ra = 0$ (right).
In the left panel, we show the logarithm of the scalar potential for the visualization purpose. 
We use the same model parameters as in Fig.~\ref{fig1}.
For comparison, the case with $B=0$ is also shown by the dashed (blue) line in the right panel.
}
\label{fig2}
\end{center}
\end{figure}

Expanding the inflaton potential around $\phi_{\rm max}$, we obtain
\bea
V_{\rm axion}(\hat{\phi}) \;\simeq\; V_0 - \lambda \hat{\phi}^4 + \cdots
\eea
with 
\bea
V_0 &=& {\cal O}(A W_0),\\
\lambda &=& \frac{W_0}{4} \left(\frac{B}{f_2^4} - \frac{A}{f_1^4} \right)
,
\eea
where  we have defined $\hat{\phi} \equiv \phi - \pi f_1$,  the constant term $V_0$ is fixed so that the potential vanishes at the  minimum,
and  the dots represent the higher order terms.
Therefore, for the parameters satisfying (\ref{cond3}), (\ref{sol1}), and (\ref{sol2}), the axion potential is equivalent 
to the hilltop quartic inflation.
In Fig.~\ref{fig2} we show the scalar potential for the saxion and the axion and its section along $\la \sigma \ra = 0$.
We can see that the saxion is stabilized during inflation and that the axion potential is given by a flat-top potential.

%
%

For sub-Planckian decay constants, the quartic coupling $\lambda$ is fixed by the Planck normalization on the curvature perturbation as
\bea
\lambda_{\rm Planck} \simeq 6.5 \times 10^{-14} \lrfp{N_e}{50}{-3},
\eea
where $N_e$ is the e-folding number.  For instance, 
the Planck normalization is satisfied if
\bea
A W_0& \simeq & 8.7 \times 10^{-14}\, f_1^4 \lrfp{N_e}{50}{-3},
\eea
for $f_1 = 2 f_2$ (i.e. $\theta = 0$ and $A=4B$). For this choice of the parameters,
the axion mass at the potential minimum is given by
\bea
\label{masscase1}
m_{\phi} &\simeq & \frac{2\sqrt{3 A W_0}}{f_1} \simeq 2.5 \times \GEV{11} \lrfp{N_e}{50}{-\frac{3}{2}} \lrf{f_1}{0.1}.
\eea
Assuming the axion coupling with the standard model (SM) gauge bosons ${\cal L} \supset c (\phi/f_1) F_{\mu \nu}\tilde{F}^{\mu \nu}$,
the decay rate is given by
\beq
\Gamma(\phi \to A_\mu A_\mu) \;=\; N_g \frac{c^2 m_\phi^3 }{4 \pi f_1^2},
\eeq
where $N_g = 8+3+1$ counts the number of SM gauge bosons.
The reheating temperature after the inflation is then estimated as
\bea
T_R & \equiv & \lrfp{\pi^2 g_*}{90}{-\frac{1}{4}} \sqrt{\Gamma} \simeq
4 \times 10^{8} \, c  \lrfp{N_e}{50}{-\frac{9}{4}} \lrfp{f_1}{0.1}{\frac{1}{2}}  {\rm GeV},
\eea
where $g_*$ counts the relativistic degrees of freedom in plasma and we have substituted $g_* = 106.75$
in the second equality.

\begin{figure}[t!]
\begin{center}
\includegraphics[width=7.9cm]{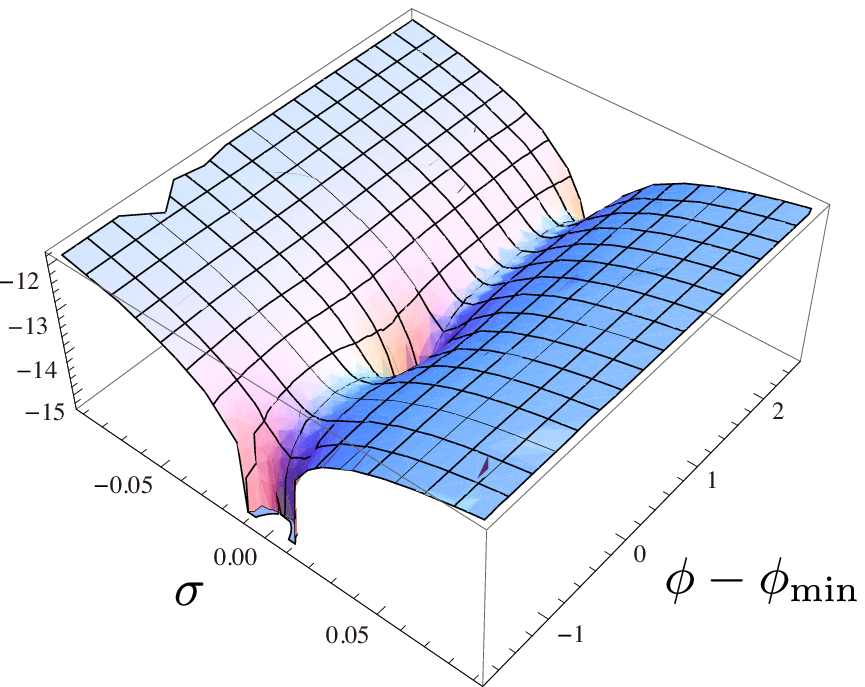}
\includegraphics[width=7.9cm]{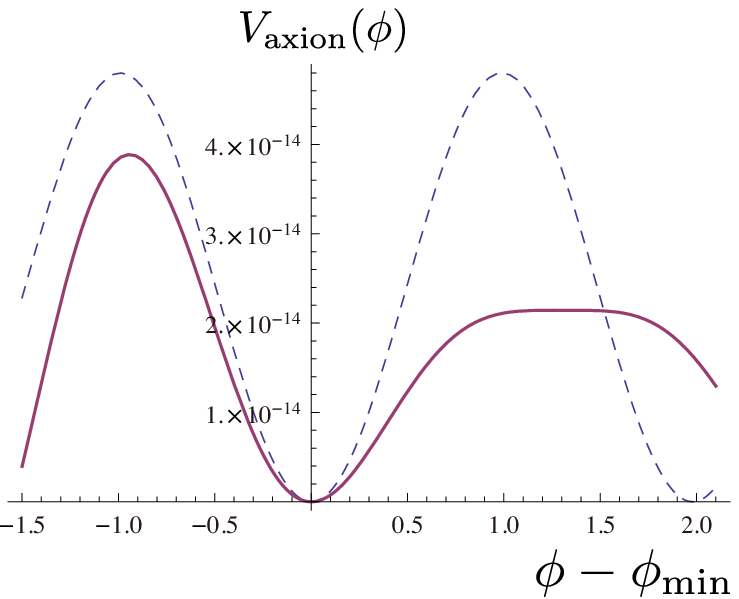}
\caption{The scalar potential for the saxion and the axion (left) and the axion potential 
at the section of $\la \sigma \ra = 0$ (right) similarly to Fig.\ref{fig2}.
We use $A=4.0\times 10^{-11},~ B=(6/7)^2 A,~a=\pi/7,~b=\pi/6,~f=0.1,~W_0=10^{-4}$ and $\theta=-7\pi/6$.
The case with $B=0$ is also shown by the dashed (blue) line in the right panel, 
where the minima are chosen to 
coincide for visualization purposes. 
}
\label{Vinf_f1=f2_2}
\end{center}
\end{figure}

So far we have adopted the special case of $f_1 = 2f_2$. The typical scales of the inflaton mass and the reheating temperature
are similar for other choices. Here let us take another case.
If the two decay constants are very close to each other, i.e., $(f_1 - f_2)/f_2 \ll 1$, we can approximate the inflaton potential by keeping the leading order term in
$(f_1-f_2)/f_2 $:
\bea
V_{\rm axion}(\phi) &\simeq&6AW_0 \lrf{f_1-f_2}{f_2} \left(v_0- 2 \cos \frac{\phi}{f_1} + \left(\pi - \frac{\phi}{f_1} \right) \sin \frac{\phi}{f_1} \right)+\cdots,
\eea
where the dots represent higher order terms of  ${\cal O}((f_1-f_2)^2/f_2^2)$, $v_0 \approx 4.8206$, and the potential maximum and minimum are located at $\phi_{\rm max}/f_1 = \pi$ and $\phi_{\rm min}/f_1 \approx
-1.3518$, respectively. This approximation is valid only for $|\phi/f_1| \ll \frac{f_2}{f_1-f_2}$. 
The potential height $V_0$, the quartic coupling $\lambda$, and the inflaton mass at the minimum are approximately given by
\bea
V_0 &\approx& 40.9 A W_0 \lrf{f_1-f_2}{f_2},\\
\lambda &\approx & \frac{AW_0}{2 f_1^4 } \lrf{f_1-f_2}{f_2},  \\
m_\phi &\approx & 5.13 \frac{\sqrt{A W_0}}{f_1}  \sqrt{\frac{f_1-f_2}{f_2}} \simeq 7.3 \sqrt{\lambda} f_1.
\label{f1=f2normal}
\eea
We can see that the inflaton mass is of similar order to Eq.~(\ref{masscase1}).
As the Planck normalization fixes $\lambda$, $AW_0$ scales as $f_2/(f_1-f_2)$, while
the inflaton potential shape itself is not significantly changed even when $f_1 \approx f_2$. Indeed, using the Planck-normalized quartic coupling,
we can rewrite the inflaton potential as
\bea
V_{\rm axion}(\phi) &\simeq& 12 \lambda_{\rm {Planck}} \,f_1^4 
\left(v_0- 2 \cos \frac{\phi}{f_1} + \left(\pi - \frac{\phi}{f_1} \right) \sin \frac{\phi}{f_1} \right)
,
\label{Vf1=f2}
\eea
in the limit of $f_1 \approx f_2$.  
The inflaton potential is shown in Fig.~\ref{Vinf_f1=f2_2}. 
Note that there are many other potential minima and maxima;
the inflation takes place near the hilltop around $\phi/f_1 \lesssim \pi$.

The spectral index for hilltop quartic inflation is predicted to be
\bea
n_s \simeq 1 - \frac{3}{N_e} = 0.94 - 0.95 ,
\eea
for $N_e = 50 - 60$.
As is well known, the predicted spectral index tends to be too low to fit the Planck result (\ref{planck}).
In the context of new inflation in supergravity~\cite{Kumekawa:1994gx,Izawa:1996dv}, the resolution
of the tension was discussed in detail in the literature, 
and it is known that the prediction of $n_s$ can be increased to be consistent with the Planck data either by
adding a logarithmic correction~\cite{Nakayama:2011ri,Bose:2013kya} or a linear term~\cite{Takahashi:2013cxa}, or by
considering higher powers of the inflaton coupling~\cite{Harigaya:2013pla}. As we shall see below, in the axion hilltop inflation, we can easily increase the spectral index by varying the relative phase $\theta$ around (\ref{sol2}).

Let us study the axion potential by varying the parameters around the solutions  (\ref{cond3}), (\ref{sol1}), and (\ref{sol2}).
Expanding the potential in terms of ${\hat \phi} = \phi - \pi f_1$, we obtain
\bea
V_{\rm axion}({\hat \phi})
&=& V_0 + \frac{6BW_0  \sin\Theta}{f_2} \,{\hat \phi} - 3 W_0 \left( \frac{A}{f_1^2} - \frac{B}{f_2^2} \cos\Theta \right) {\hat \phi}^2
-\frac{BW_0 \sin \Theta}{f_2^3} \, {\hat \phi}^{3}\non\\
&& - \frac{1}{4}W_0 \left( \frac{B}{f_2^4} \cos \Theta - \frac{A}{f_1^4} \right) {\hat \phi}^4 + \cdots,
\eea
where we have defined $\Theta \equiv \theta + \pi f_1/f_2$.  It is the linear term in $\hat \phi$ that affects the inflaton dynamics significantly.
As pointed out in Ref.~\cite{Takahashi:2013cxa}, if there is a small linear term in the hilltop quartic inflation model,  
the inflaton field value at the horizon exit of cosmological scales 
can be closer to the hilltop, making the curvature of the potential smaller and therefore increasing the spectral index.

\begin{figure}[t!]
\begin{center}
\includegraphics[width=7.9cm]{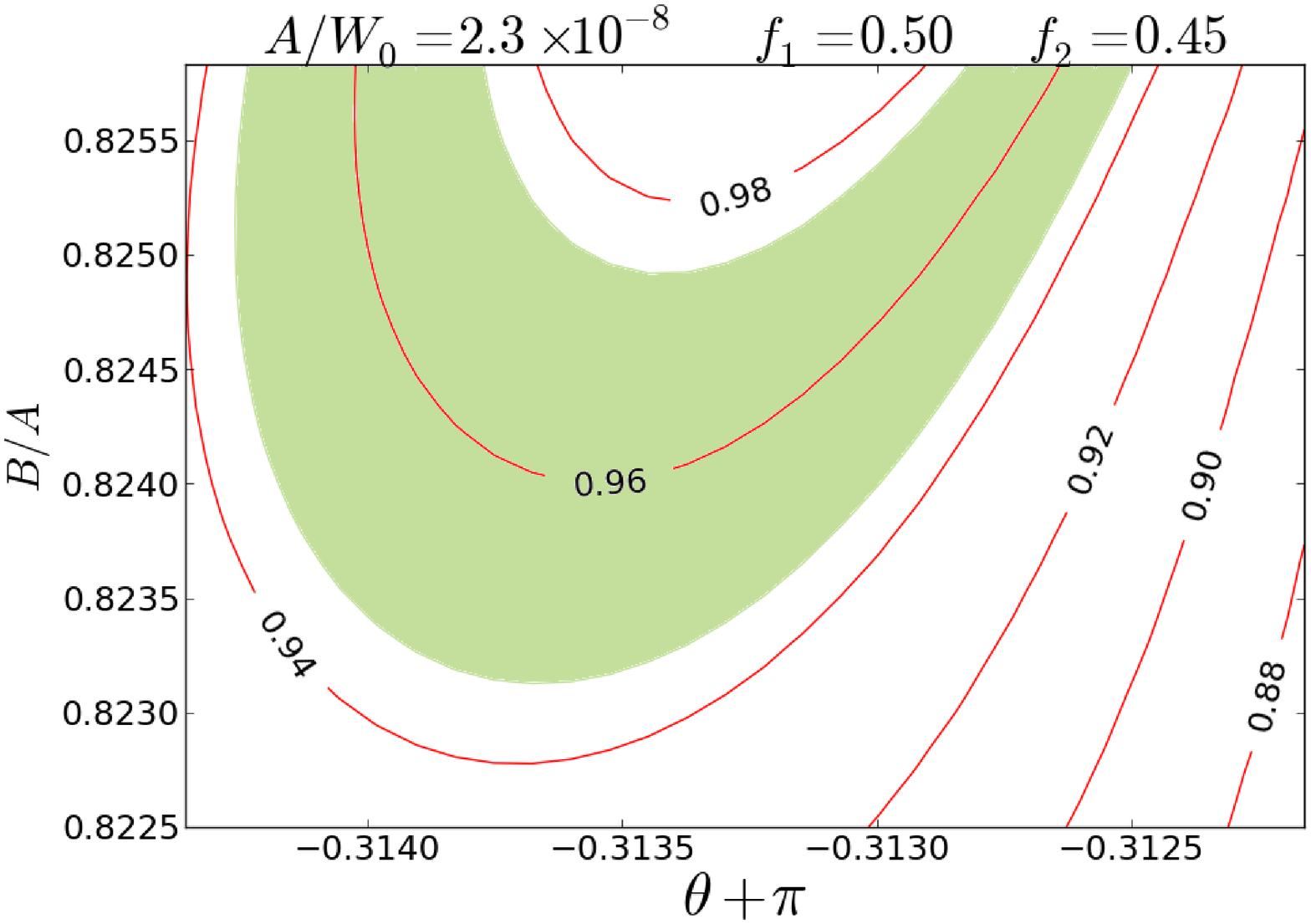}
\includegraphics[width=7.9cm]{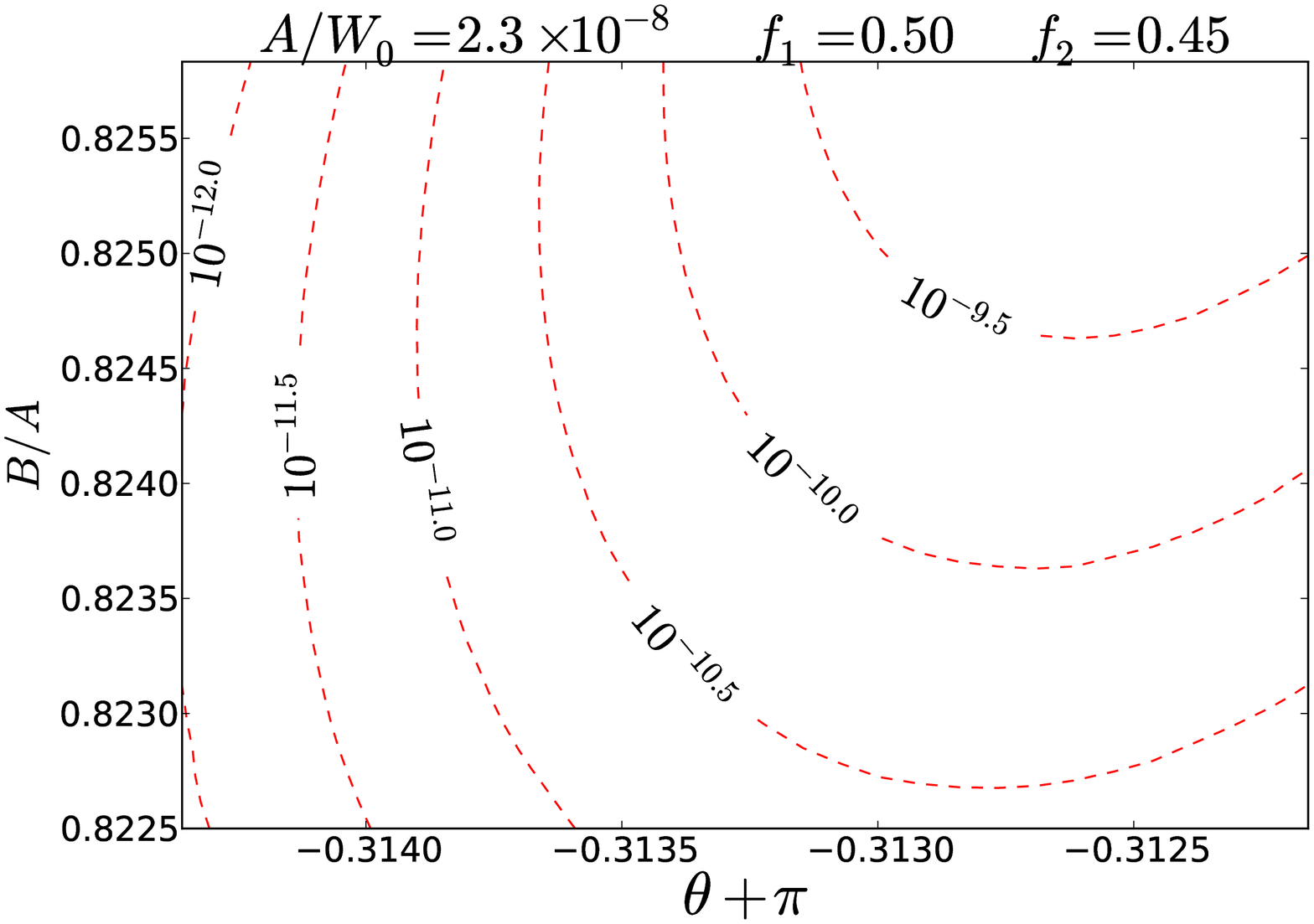}
\caption{Plots of $n_s$ (left) and $r$ (right) for varying values of $B$ and $\theta$ for fixed decay 
constants $f_1 = 0.5$ and $f_2 = 0.45$, which corresponds to the case of  $f_1 \approx f_2$ studied in the text. 
The green shaded region corresponds to the 2$\sigma$ allowed region
for $n_s$ from the Planck data.}
\label{fig3}
\end{center}
\end{figure}

We have numerically solved the inflaton dynamics based on the potential given by  \EQ{pot}
to evaluate the predicted values of $n_s$ and $r$. 
To be concrete, we have varied the model parameters $B/A$ and $\theta$ 
around the solutions  (\ref{cond3}), (\ref{sol1}), and (\ref{sol2}) with
$f_1 = 0.5$ and $f_2 = 0.45$. The results are shown in  Fig.~\ref{fig3}.
Note that the Planck normalization can be satisfied by varying 
$W_0$ for fixed $A/W_0$ and $B/W_0$ without affecting the predicted values of $n_s$ and $r$.
From Fig.~\ref{fig3} we can see that the spectral index can be increased to fit the 2$\sigma$ limit of the Planck data shown
by the shaded (green) region. We have also confirmed that the spectral index can be similarly increased to give a good fit to the Planck
data for different values of $f_1$ and $f_2$, e.g. $f_1 = 2 f_2$. In general, for smaller values of the decay constants, the deviation from the solution (\ref{sol2})
must be smaller. 
On the other hand, as expected, the tensor-to-scalar ratio $r$ is well below the upper bound from the Planck data ($r < 0.11$).
For a larger value of the decay constant, e.g.  $f = 1$ ($f_1 \approx 2.25$) $r$ can be as large 
as $\sim 10^{-3}$ in the allowed region of $n_s$.

\begin{figure}[t!]
\begin{center}
\includegraphics[width=7.9cm]{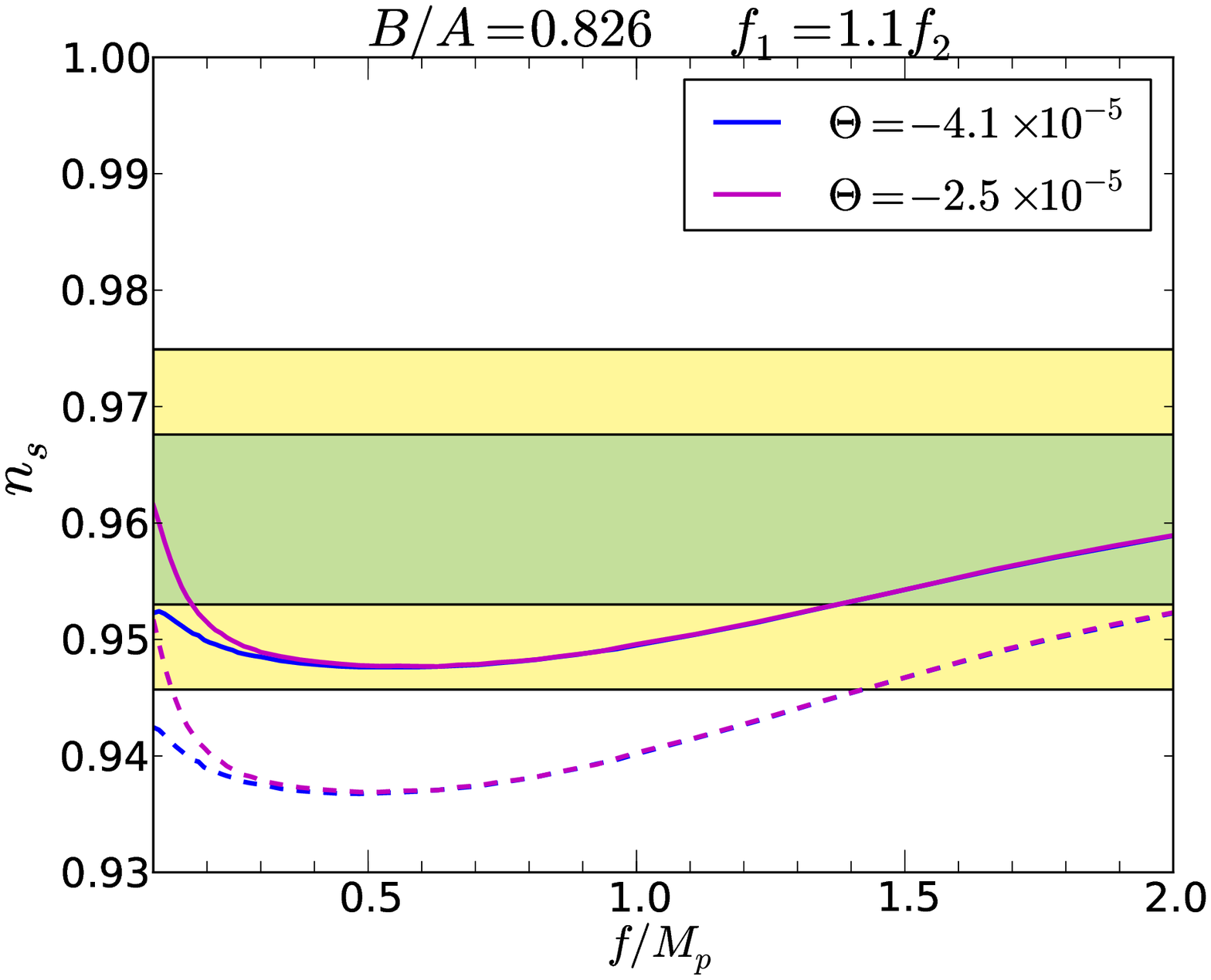}
\includegraphics[width=7.9cm]{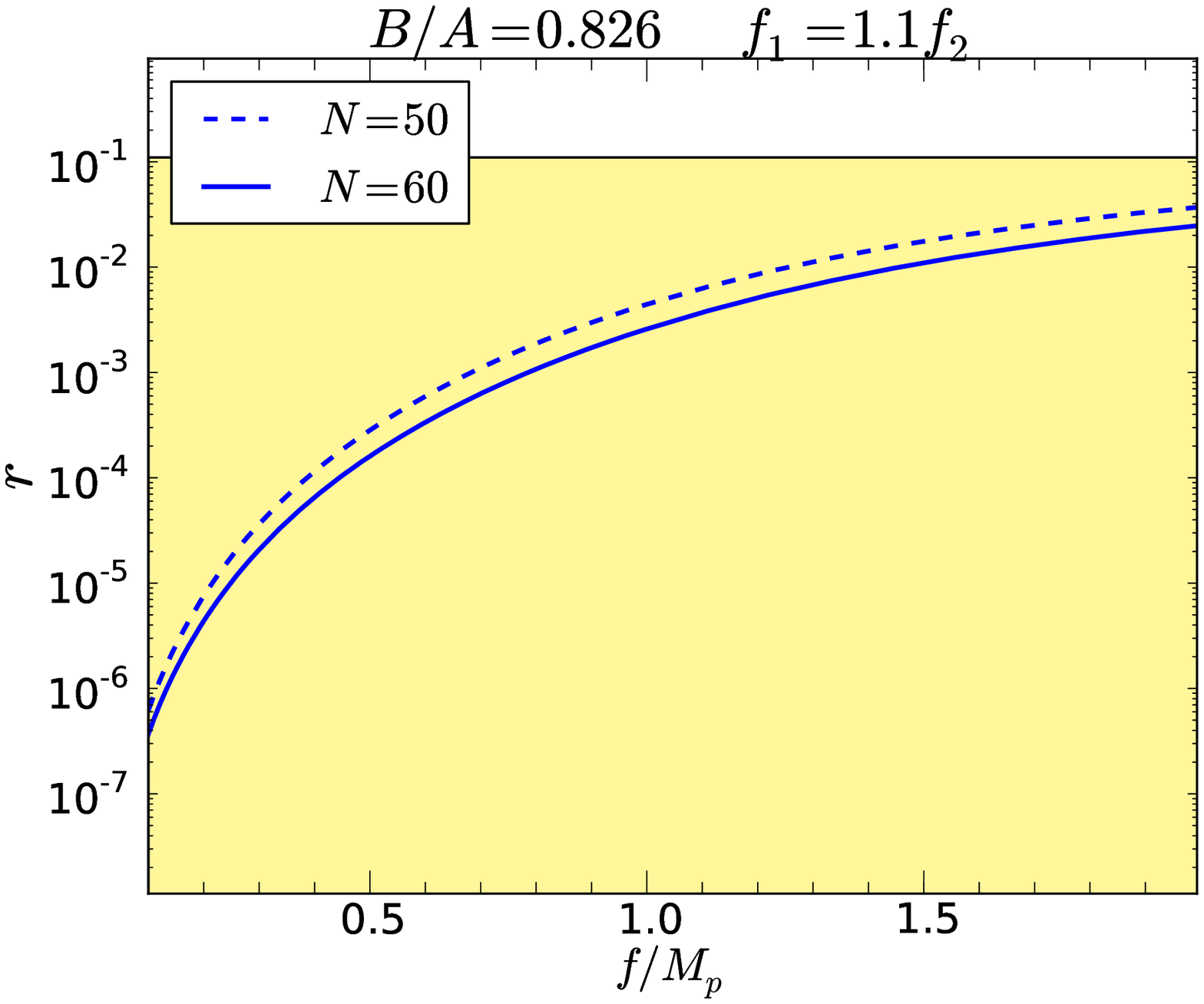}
\caption{ Plots of $n_s$ (left) and $r$ (right) as a function of $f/M_p$. In the left figure, $\Theta \equiv \theta + \pi f_1/f_2$. In the right figure, there was no significant difference in the behavior of $r$ for the two values of $\Theta$, hence we chose $\Theta = -4.1\times 10^{-5}$. Solid (dotted) lines correspond to $N_e=60$ ($N_e=50$).}
\label{fig5}
\end{center}
\end{figure}

 Fig.~\ref{fig5} also shows the behavior of $n_s$ and $r$ as a function of $f_1$ with the same parameters as in Fig.~\ref{fig2}. The behavior is 
 similar to a hilltop quartic model as discussed in Ref.~\cite{Czerny:2014wza} for $\theta = 0$, but the spectral index can be increased by allowing
 a non-zero relative phase.

\section{UV completion based on  string-inspired model}
\label{Secstring}

\subsection{Set-up}
We now provide a further UV completion of the effective SUGRA model given in the previous section,  
based on the string-inspired model.
Let us consider a model with three K\"ahler moduli on a Calabi-Yau space with the following K\"ahler
and super-potentials\,\footnote{
A similar UV completion may be possible in a LARGE volume scenario~\cite{Balasubramanian:2005zx}
with string-loop corrections and non-perturbative superpotentials,  if a (moderately) big cycle allows
a gauge coupling which generates the axion mass through non-perturbative effects. 
A large mass hierarchy between the saxion and the axion can then be realized: the saxion mass is 
suppressed by the power of the Calabi-Yau volume while the axion mass is exponentially suppressed by the volume~\cite{Cicoli:2014sva}.  
} 
\bea
K &=& -2\log(t_0^{3/2} - t_1^{3/2}-t_2^{3/2});~~~t_i = (T_i +T_i^{\dag}) ~~~{\rm for}~i=0,1,2, \\
W &=& W_0 - C e^{-\frac{2\pi}{N}T_0} - D e^{-\frac{2\pi}{M}(T_1+T_2)} 
+ A e^{-\frac{2\pi}{n_1}T_2}+ B e^{-\frac{2\pi}{n_2}T_2},
\label{siW}
\eea
where $T_i$ are complex K\"ahler moduli, and
$W_0,~A,~B,~C$ and $D$ are determined by
the vacuum expectation values (VEVs) of heavy dilaton/complex structure stabilized via three-form flux 
compactification~\cite{Grana:2005jc,Blumenhagen:2006ci}. (See \cite{Giryavets:2003vd} for  realization of a small $W_0$.)
The exponential terms in the superpotential are assumed to be generated by gaugino condensations in a 
pure $SU(N)\times SU(M)\times SU(n_1)\times SU(n_2)$ gauge theory. Those gauge fields are living on the D-branes 
wrapping on the divisors whose volume is determined by the real part of the moduli, 
$T_0$, $T_1+T_2$, $T_2$ and $T_2$ respectively. In other words,
(at least some part of) the gauge coupling of each gauge group is given by the corresponding moduli.
We define  $T \equiv T_1 + T_2 $ and $\Phi \equiv -T_1 +T_2$, and express the lowest component of $\Phi$ as 
$\Phi =   \sigma + i \phi$ for later use. 

Using the U(1)$_R$ symmetry and an appropriate shift of the imaginary components of the moduli fields, 
we can take $W_0$, $A$, $C$, and $D$ real and positive without loss of generality. We will include a relative phase
in $B$ by replacing it with $B e^{-i \theta}$ where $B$ is a real and positive constant. 
 We assume that those parameters satisfy
\bea
&& A,  B, C, D = {\cal O}(1),~~~W_0 \ll 1.
\eea
We also assume that $N,~M,~n_1$ and $n_2$ are integers satisfying~\footnote{
We shall see that the axion hilltop inflation with $f_1 \approx f_2$ 
is realized for this choice of the parameters. The other cases such as $f_1 = 2 f_2$ can also be
 realized if there is a hierarchy between $A$ and $B$. See Appendix \ref{Sectuning}.
}
\bea
&& n_1 \sim n_2~~{\rm and}~~n_1\ne n_2\\
&& 2n_1 < M \lesssim N.
\label{n1MN}
\eea 
The mild hierarchy between $(M,N)$ and $(n_1,n_2)$ implies that $T_0$ and $T=T_1+T_2$ are stabilized in a
supersymmetric manner by the first two exponentials. On the other hand $\Phi = -T_1 + T_2$ remains relatively
light, and this combination becomes the axion supermultiplet  in the previous section.
We shall see that, while $\sigma = {\rm Re}[\Phi]$ can be stabilized by the SUSY breaking effect through 
the K\"ahler potential, the axion,  $\phi = {\rm Im}[\Phi]$,   acquires an even lighter mass by the last two exponentials. 
In order to have successful inflation with sub-Planckian decay constants, the resultant two exponentials 
expressed in terms of $\Phi$ must be comparable in size. This is possible for $A \sim B$, if $n_1$ is close to $n_2$
within $10\%$ or so.\footnote{
Precisely speaking, this is the case if $|n_2 - n_1| \lesssim n_1 n_2/\pi \la T \ra $.
If $n_1$ is not close to $n_2$, some hierarchy between $A$ and $B$ is necessary.}

\subsection{Heavy moduli stabilization}
We first study the stabilization of the heavy moduli, $T_0$,  $T$ and $\sigma$.
For $A=B=0$, the model is reduced to the string-theoretic QCD axion model considered in Ref.~\cite{Choi:2006za};
there exists a Minkowski vacuum where, while the other moduli are stabilized,  ${\rm Im}[\Phi]$ remains (almost) 
massless and eventually becomes the QCD axion. Because of the assumed mild hierarchy between
$(M,N)$ and $(n_1, n_2)$, our model is similar to this model as long as the heavy moduli stabilization is concerned. 

The scalar potential of the moduli and  the sequestered SUSY-breaking up-lifting potential $V_{\rm up}$ are given by
\bea
V = V_{\rm moduli} + V_{\rm up},~~~{\rm where}~~
V_{\rm up} = \hat{\epsilon}\, e^{2K/3};~~~\hat{\epsilon} ={\cal O}(W_0^2),
\eea
where $V_{\rm moduli}$ is given by \EQ{sugra} with the above K\"ahler and super-potentials, and
$\hat{\epsilon}$ is fixed so that a (nearly) Minkowski vacuum is realized in the low energy. 
The moduli stabilization is determined by the
conditions for extremizing the potential $V$,  $D_{T_0}W \simeq D_{T}W \simeq \partial_{\Phi}K \simeq 0$:
\begin{align}
 \frac{2\pi}{N}T_0 &\simeq \frac{2\pi}{M} T \simeq \log\bigg[\log(1/W_0)/W_0\bigg] \gg 1,~~
{\rm Re}[\Phi]= 0.
\label{SolofT}
\end{align}
The VEVs of these moduli fields fix the volume of the Calabi-Yau space ${\cal V}$ as well as the gravitino mass as
\bea
{\cal V} & = & t_0^{3/2} - \frac{t^{3/2}}{\sqrt{2}},\\
m_{3/2} &=& W_0/{\cal V},
\eea
where $t = T+T^{\dag}$. The moduli masses are given by
\begin{align}
m_{T_0}& \simeq m_{T} \simeq \log(M_{\rm Pl}/m_{3/2}) m_{3/2},\\
m_{\sigma} &\simeq \sqrt{2} m_{3/2},
\end{align}
where we used a fact that the up-lifting potential after integrating out $T_0$ and $T$ is approximately given by
\bea
V_{{\rm up}L} &\approx & 3m_{3/2}^2 + f^2 m_{3/2}^2 (\Phi +\Phi^{\dag})^2 + \cdots.
\eea
See Appendix~\ref{higherphi} for the higher order terms in $V_{{\rm up}L}$. 
Note that the axion $\phi$ remains massless in this case, which should be contrasted
to the original KKLT \cite{Kachru:2003aw}. 
The SUSY-breaking $F$-terms of the moduli fields are given by
\bea
\frac{F^{T_{i}}}{T_i+T_i} \sim 
\frac{m_{3/2}}{\log(M_{\rm Pl}/m_{3/2})}  \sim m_{\rm soft} \lesssim m_{3/2},
\eea
and, therefore, all the SUSY particles generically acquire a soft SUSY breaking 
 slightly  lighter than the gravitino mass through the modulus mediation. The anomaly mediation also gives
a comparable contribution to the soft mass.
Although the mass of the SUSY SM particles are relevant for the observed SM-like Higgs boson mass,
they are irrelevant for the inflaton dynamics during inflation.

For $A,~B \neq 0$, the axion can have a non-zero mass much smaller than the gravitino mass\,\footnote{
In order to implement the QCD axion, one would need to introduce another moduli field.
Alternatively, the QCD axion may originate from an open string mode.
}, while  the stabilization of the heavy moduli is not changed drastically.
This is because, as long as (\ref{n1MN}) is satisfied,  the last two exponential terms in the superpotential (\ref{siW}) 
are much smaller than the others: $ A e^{-\frac{2\pi}{n_1}\langle T_2 \rangle} 
\sim W_0^{\frac{M}{2n_1}} \ll  W_0 (\ll 1)$.

\subsection{Axion inflation in low energy effective theory}
\label{SECLET}
In this subsection, we focus on the lightest axion multiplet $\Phi$
at  scales below the heavy  moduli masses. We discuss multi-natural inflation within
the low energy effective theory of this string-inspired model, using the results obtained in the previous section.

After integrating out the heavy moduli $T_0$ and $T $,
we obtain the low energy effective theory for $\Phi$,
\bea
\label{KL}
K_L & \approx & 
\frac{f^2}{2} (\Phi +\Phi^{\dag})^2 
+ \cdots , \\
\label{WL}
W_L & \approx & W_0 + \hat{A} e^{-\frac{\pi}{n_1}\Phi }+ \hat{B} e^{-\frac{\pi}{n_2}\Phi - i \theta}.
\eea
The higher order terms in $K_L$  are given in Appendix \ref{higherphi}. Here, we have defined
\bea
f^2 &\equiv & \frac{3}{2 \sqrt{2} \sqrt{t} {\cal V}} \lesssim 1, ~~~
\hat{A}  \equiv  A e^{-\frac{\pi}{n_1}\langle T \rangle},~~~ \hat{B} \equiv B e^{-\frac{\pi}{n_2}\langle T \rangle}.
\eea
A natural value of $f$ is considered to be of order $0.1$ since it is on the order of the string scale.

The above K\"ahler and super potentials are equivalent to (\ref{W1}) and (\ref{K2}) studied in the previous section, 
and successful axion inflation is possible for a certain choice of the parameters. The  parameters are related as
\bea
&&a =  \frac{\pi}{n_1},~~~~~~~b = \frac{\pi}{n_2}\\
&&f_1 = n_1 f_a,~~f_2 = n_2 f_a, 
\eea
where we have defined $f_a \equiv \sqrt{2}f / \pi$.

Note that the prefactors  of the exponentials, ${\hat A}$ and ${\hat B}$, are comparable to each other, and
much smaller than $W_0$,
\bea
\hat{A} \sim \hat{B} \sim W_0^{M/2n_1} \ll W_0 (\ll 1).
\eea
As we have seen before, this hierarchy is one of the essential ingredients for multi-natural inflation.
Indeed,  the  ratio of the axion mass to the saxion mass is much smaller than unity;
\bea
\frac{m_{\phi}^2}{m_{\sigma}^2} \sim \frac{\hat{A}W_0}{W_0^2} \sim W_0^{\frac{M}{2n_1} -1} \ll 10^{-2}~~~{\rm for}~
 M > 2n_1,
\eea
 and therefore the saxion remains stabilized near the origin, $\la \sigma \ra \simeq 0$.

In order to have axion hilltop inflation, the model parameters must satisfy
the relations  (\ref{cond3}), (\ref{sol1}), and (\ref{sol2}) to a high accuracy.
In particular, the condition  (\ref{cond3}) reads
\bea
\hat{A} \approx \bigg(\frac{n_1}{n_2}\bigg)^2\hat{B},
\eea
which implies that the gaugino condensation from $SU(n_1)$ should be comparable 
to that from $SU(n_2)$ in size. In terms of the gauge couplings at the cut-off scale,
this condition can be expressed as
\bea
\frac{g_2^2}{g_1^2} \sim \frac{n_1}{n_2}\bigg[
1 + \frac{n_2 g_2^2}{8\pi^2}\log\left(
\frac{n_2}{n_1}
\right)
\bigg],
\eea
where $g_i~(i=1,2)$ is the gauge coupling in $SU(n_i)$ gauge group at the cut-off scale
and we have used the fact that $\hat{A}$ and $\hat{B}$ are proportional to $n_1$ and $n_2$, respectively:
$W_{SU(n_i)} = n_i \Lambda_{SU(n_i)}^3 = n_i e^{-8\pi^2/n_i g_i^2}$, where the $\theta$-term is omitted.
Thus,  successful axion hilltop inflation requires a certain relation between the rank of the gauge groups 
and the value of the  gauge couplings of  a  gauge theory where gaugino condensations form in the low energy.
This is equivalent to the relation between the world volume of the relevant D-branes and the number of such branes
in string theory. For further discussions on the magnitude of $A$ and $B$, see Appendix \ref{Sectuning}.

Lastly let us express the inflaton mass and the Hubble parameter during inflation in terms of 
the gravitino mass. To this end we consider the case where $n_1$ is not degenerate with $n_2$. \footnote{
When $n_1$ is close to $n_2$, the factor proportional to $(n_1-n_2)/n_2$ should be included as 
discussed in the previous section.}
The axion mass at the potential minimum, the potential height and the inflation scale are estimated as
\bea
m_{\phi}^2 &\sim & \frac{\hat{A} W_0}{n_1^2 f_a^2} 
\sim m_{3/2}^2 \bigg(\frac{m_{3/2}}{n_1 f_a}\bigg)^{2}  
\bigg(\frac{m_{3/2}}{M_{\rm Pl}}\bigg)^{\frac{M}{2n_1}-3}
\bigg[\frac{1}{\log(M_{\rm Pl}/m_{3/2})}\bigg]^{\frac{M}{2n_1}}
, 
\label{m_3/2_minf}
\\
V_0 &\sim & \hat{A} W_0 
\sim m_{3/2}^4 \bigg(\frac{m_{3/2}}{M_{\rm Pl}}\bigg)^{\frac{M}{2n_1}-3} 
\bigg[\frac{1}{\log(M_{\rm Pl}/m_{3/2})}\bigg]^{\frac{M}{2n_1}}
, \\ 
H_{\rm inf}  &\sim & \frac{\sqrt{V_0}}{M_{\rm Pl}} \sim
m_{3/2} \bigg(\frac{m_{3/2}}{M_{\rm Pl}}\bigg)^{\frac{M-2n_1}{4n_1}} 
\bigg[\frac{1}{\log(M_{\rm Pl}/m_{3/2})}\bigg]^{\frac{M}{4n_1}}
\eea
for $M>2n_1$.
For instance, $m_{\phi} \simeq 10^{11}$GeV is obtained for 
$m_{3/2} \simeq 10^{14}$GeV, $f_a \simeq 10^{17}$ GeV, $n_1 =6$ and $M= 24$.
The last equation implies that the Hubble parameter during inflation is necessarily smaller than the gravitino mass,
\bea
H_{\rm inf} < m_{3/2},
\eea
which enables us to avoid the moduli destabilization \cite{Kallosh:2004yh}.
This is because the flatness of the inflaton potential is  not due to SUSY, but 
(mostly) due to both the axionic shift symmetry and  the dynamical origin of the potential.

\subsection{Reheating and leptogenesis}

In order to have successful inflation, the inflaton must transfer its energy to the SM particles.
Also, as any pre-existing baryon asymmetry is diluted by the inflationary expansion, 
the right amount of baryon asymmetry must be created after inflation.
Here we study reheating and the baryon number generation through leptogenesis~\cite{Fukugita:1986hr}.

As for reheating, the axion will decay into the SM gauge bosons through its couplings to the SM gauge fields,
\bea
{\cal L}_{\rm SM} &=& \frac{1}{16\pi}\int d^2 \theta T_2 {\cal W}^{\alpha}_{\rm SM} {\cal W}_{\alpha {\rm SM}} + {\rm h.c.}
\supset \frac{1}{32\pi}\int d^2 \theta \Phi {\cal W}^{\alpha}_{\rm SM} {\cal W}_{\alpha {\rm SM}}+ {\rm h.c.} \\ 
\Gamma^{\phi}_g &\equiv &
\Gamma(\phi \to 2A_{\mu}) \simeq \frac{N_g}{32 \pi  t^2}\frac{{m_\phi }^3 }{f^2}
\simeq 
\frac{N_g g_{\rm SM}^4 }{4096 \pi ^5}\frac{m_{\phi}^3}{f_a^2},
\eea
where $t \simeq 16\pi/g_{\rm SM}^2$.
We have assumed that the SM is living on the D-brane wrapping on $T_2$-cycle.
Thus, the reheating temperature is given by
\bea
T_R^g \simeq \left(\frac{\pi^2 g_*(T_R) }{90 }\right)^{-1/4}\sqrt{\Gamma^{\phi}_g M_{\rm Pl}} \simeq 4\times 10^{5}~{\rm GeV}
\bigg(\frac{m_{\phi}}{10^{11}{\rm GeV}}\bigg)^{3/2} \bigg(\frac{f_a}{10^{17}{\rm GeV}}\bigg)^{-1},
\label{TR}
\eea
if this is the main decay mode.\footnote{
The axion cannot decay into the SM gauginos since their mass is heavier than the axion mass.
Even if it is possible, the estimation will not be changed drastically as studied in \cite{Higaki:2013lra};
the R-parity may have to be violated to avoid the overabundance of dark matter.
The moduli-induced baryogenesis~\cite{Ishiwata:2013waa} may work in this case.
} 
Here we have used $N_g=12$ and $g_{\rm SM}^2/4\pi =1/25$ and $g_*(T_R) = 106.75$.

Next, let us consider the origin of the baryon asymmetry. 
Among various baryogenesis scenarios, leptogenesis is a plausible and interesting possibility
in the light of the observed neutrino masses and mixings. 
We focus on non-thermal leptogenesis \cite{LQlepto,Asaka:1999yd},
because it can generate a sufficient amount of baryon asymmetry with a relatively low reheating temperature.

\begin{figure}[t!]
\begin{center}
\includegraphics[width=10cm]{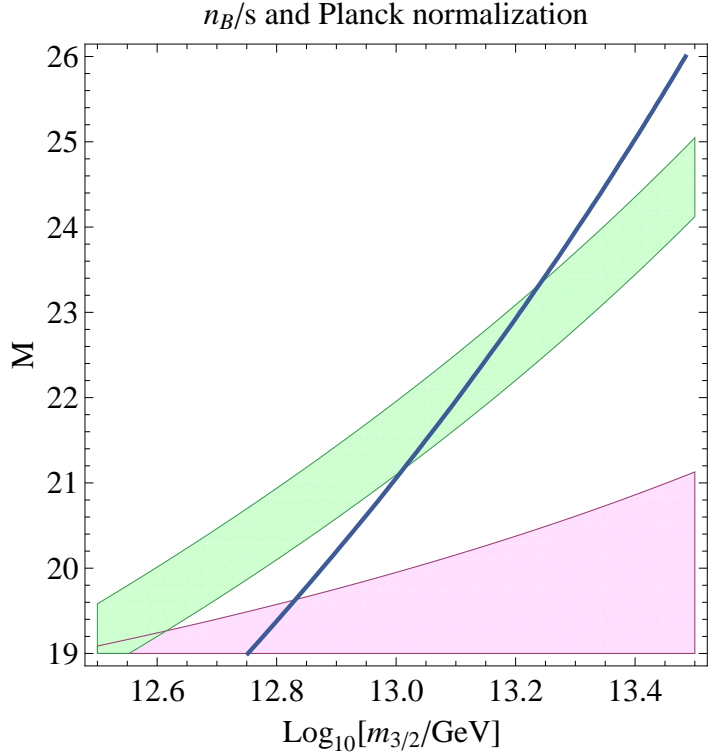}
\caption{Plots for the baryon asymmetry and the Planck normalization in the $(M,m_{3/2})$-plane.
The green shaded region shows $0.5 \times 10^{-10} \leq  n_B/s \leq 1.5 \times 10^{-10}$.
The blue line shows the Planck normalization corresponding to $\lambda_{\rm Planck} = 3.7 \times 10^{-14}$ for $N_e \simeq 60$.
In the red shaded region, the saxion vacuum deviates from the origin.
We used $n_1 = 7$, $n_2 = 6$, $f_a = 2.3\times 10^{17}$ GeV and $C_{\nu^c} = 8$; 
$f_1 = 7f_a$ and $f_2 =6f_a$.
Then $m_{\phi} \simeq 9.6 \times 10^{11}$ GeV is obtained, using Eq.(\ref{f1=f2normal}).
We find also $m_{3/2} \sim 10^{13}$ GeV, $m_{\nu^c} \sim 10^{11}$ GeV and $T_R \sim 10^{7}$ GeV
in the viable region, where we can obtain the correct curvature perturbation and baryon asymmetry.}
\label{FigLepto}
\end{center}
\end{figure}

The right-handed neutrino ${\nu}^c$ can be produced by the axion decays 
via the coupling below
\bea
W = C_{{\nu}^c} e^{-\frac{2\pi}{n_1}T_2} \nu^c \nu^c ,
\eea
in which $C_{\nu^c}$ is a constant. 
Note that the mass of the neutrinos is given by
\bea
m_{\nu^c} \simeq 2 C_{{\nu}^c} M_{\rm Pl}\bigg[\frac{m_{3/2}/M_{\rm Pl}}{\log(M_{\rm Pl}/m_{3/2})}\bigg]^{\frac{M}{2n_1}},
\label{m_3/2_mnu}
\eea
while $m_{\phi} \sim W_0^{(M+2n_1)/4n_1}/f$; 
$m_{{\nu}^c}/m_{\phi} \sim W_0^{\frac{(M-2n_1)}{4n_1}}f < 1$.
For instance, one obtains $m_{{\nu}^c} \simeq 10^{10}$ GeV when taking $W_0 = 10^{-4}$, $M=24$, $n_1=6$ and $C_{\nu^c}= 30$.
Such a term is generated when the right-handed neutrino is coupled to the gauge field of $SU(n_1)$:
$ \int d^2 \theta {\nu}^c {\nu}^c [{\cal W}^{\alpha}{\cal W}_{\alpha}]_{SU(n_1)}$.
(A similar origin is also discussed in the literatures \cite{Blumenhagen:2006xt,Ibanez:2006da,Blumenhagen:2009qh}.)
The decay fraction is given by
\bea
\Gamma_{{\nu}^c}^{\phi} 
\equiv
\Gamma (\phi \to 2{\nu}^c) \simeq \frac{1}{16\pi}\left(\frac{m_{{\nu}^c}}{n_1 f_a}\right)^2 m_{\phi}.
\eea
The reheating will proceed mainly via the decay into the neutrinos if 
\bea
\frac{m_{\nu^c}}{m_{\phi}} \gtrsim 10^{-2}.
\eea
The reheating temperature is then estimated as
\bea
T_R^{\nu^c} \simeq 1 \times 10^7{\rm GeV}\left( \frac{m_{\phi}}{10^{11}~{\rm GeV}}\right)^{1/2}
\left( \frac{m_{\nu^c}}{10^{10}~{\rm GeV}}\right)
\left( \frac{n_1 f_a}{10^{17}~{\rm GeV}}\right)^{-1} .
\eea
Using the reheating temperature $T_R \simeq \sqrt{(\Gamma^\phi_{{\nu}^c} + \Gamma^\phi_g)  M_{\rm Pl}}$, the net baryon asymmetry is written as
\bea
\frac{n_B}{s} \simeq \frac{28}{78}\cdot \epsilon \cdot \frac{3}{2}  \frac{T_R}{m_{\phi}}B_{\nu^c}^{\phi},
\eea
with
\bea
\epsilon \simeq \frac{3}{16\pi}\frac{m_{\nu_3}m_{\nu^c}}{v^2}\delta_{\rm eff},
\eea
where $B_{\nu^c}^{\phi}$ is the decay fraction into the right-handed neutrino,
 $m_{\nu_3}$ is the heaviest neutrino mass, $v\simeq 174$ GeV is the Higgs VEV, 
and $\delta_{\rm eff}$ is an effective CP-phase in the neutrino Yukawa couplings.
Here  we have assumed the SM contribution for the sphaleron process and
assumed that the axion decays mainly into lightest right-handed neutrino.
Then, we can generate a right amount of the baryon asymmetry,
\bea
\frac{n_B}{s} \simeq 5.4\times 10^{-11}\delta_{\rm eff} 
\left(\frac{T_R/m_{\phi}}{10^{-4}}\right)\left(\frac{m_{\nu_3}}{0.05{\rm eV}}\right)
\left( \frac{m_{\nu^c}}{10^{10}~{\rm GeV}}\right)
\left( \frac{B_{\nu^c}^{\phi}}{1}\right).
\eea
Thus, if the axion decays mainly into the lightest right-handed neutrinos, non-thermal leptogenesis
works successfully. 
In Fig.\ref{FigLepto}, the plots for the baryon asymmetry are shown with $\delta_{\rm eff} = 1$, 
using the relation between the mass scales and the gravitino mass of Eq.(\ref{m_3/2_minf}) and (\ref{m_3/2_mnu}).
In the successful case, we can obtain $m_{3/2} \sim 10^{13}$ GeV, $m_{\nu^c} \sim 10^{11}$ GeV and $T_R \sim 10^{7}$ GeV.

No dark matter candidate has been considered in the setup so far. As the inflation scale is lower than the
typical scale of the soft SUSY breaking mass, no SUSY particles are produced during and after reheating.
Therefore,  the QCD axion or  light axion-like particles,  light sterile neutrinos, hidden photons, etc., or
their combination, are candidates for dark matter.

Finally, we give a comment on the SM-like Higgs boson mass.
In this model, the soft SUSY-breaking  terms will be on the order of the gravitino mass or slightly smaller \cite{Choi:2005ge}.
For $m_{\rm soft} \sim m_{3/2}/\log(M_{\rm Pl}/m_{3/2}) \sim 10^{12}$ GeV in the viable region for inflation, 
the Higgs mass becomes $\sim 126$ GeV for $\tan \beta \sim 1$ \cite{Giudice:2011cg}.
The small value of $\tan \beta$ can be realized in the presence of a shift symmetry or an exchange symmetry in the Higgs 
sector~\cite{Hebecker:2012qp,Ibanez:2012zg,Hebecker:2013lha}.

\section{Discussion and Conclusions}
\label{Secsum}

We have studied multi-natural inflation~\cite{Czerny:2014wza} in SUGRA for a UV completion.
In this model the inflaton potential mainly consists of two sinusoidal functions that are comparable in size, but
have different periodicity.  For sub-Planckian values of the decay  constants,  this model is reduced 
to the hilltop quartic inflation in a certain limit. It is known however that the predicted spectral index, $n_s \simeq 0.94$,
for the e-folding number $N_e \simeq 50$ tends to be too low to explain the Placnk results. 
We have shown that, allowing a relative phase between 
the two sinusoidal functions, the spectral index can be increased to give a better fit to the Planck data
based on the axion hilltop inflation in SUGRA.
We have also considered a further UV completion based on a string-inspired framework, and have shown 
that the  axion hilltop inflation model can be indeed obtained in the low energy limit.

The axion hilltop inflation requires a rather flat potential near the (local)  potential maximum.
For realizing the flat-top potential, 
 there should exist a relation between the ratio of the decay constants 
and the dynamical scale: $(f_1/f_2)^2 \approx A/B$.
This in turn implies that in string theory there should be a relation between the world volume of D-branes where the non-perturbative 
effects occur and the number of such branes.

It is also noted that because we have used supergravity, the gravitino mass is related to physical quantities. 
For successful inflation, the typical scale of the gravitno mass is $m_{3/2} \sim \GEV{13}$, whereas the soft mass
is about one order of magnitude smaller, $m_{\rm soft} = \GEV{12}$, while the inflaton mass is $m_\phi \sim \GEV{11}$.
Thus, the SUSY particles are not produced in the Universe after reheating. 
The dark matter candidates can be considered such as the QCD axion, axion-like particles, or sterile neutrinos, if they exist.
In particular, a light dark matter is interesting in light of its longevity.
The recently discovered X-ray line at $3.5$\,keV~\cite{Bulbul:2014sua,Boyarsky:2014jta}
 may be due to the decay of one of such light dark matter\footnote{See the recent works on explaining
 the $3.5$\,keV X-ray line by axions~\cite{Higaki:2014zua} or sterile neutrinos~\cite{Ishida:2014dlp}.}.
It may be possible to explain the baryon asymmetry from the inflaton decay into right-handed neutrinos.

\vspace{5mm}
{\it Note added:}
After the submission of our paper, the BICEP2 experiment found the primordial B-mode
polarization~\cite{Ade:2014xna}, which suggests $r = 0.20^{+0.07}_{-0.05}$.
 Although we have focused on the hilltop inflation limit of
the multi-natural inflation when we evaluate the spectral index and the tensor-to-scalar ratio, 
most of the discussion including the realization of the multi-natural inflation in supergravity and
string-inspired set-up, the reheating, and leptogenesis is applicable to a more general 
multi-natural inflation. In particular, for such a large value of $r$, the inflaton mass will be of order
$\GEV{13}$, leading to the reheating temperature close to $\GEV{9}$ (cf. \EQ{TR}). Therefore,
thermal leptogenesis will be possible. See also the related papers on the multi-natural 
inflation~\cite{Czerny:2014qqa, Czerny:2014wua} that appeared after BICEP2.

\section*{Acknowledgment}
This work was supported by Grant-in-Aid for  Scientific Research on Innovative
Areas (No.24111702, No. 21111006, and No.23104008) [FT], Scientific Research (A)
(No. 22244030 and No.21244033) [FT], and JSPS Grant-in-Aid for Young Scientists (B)
(No. 24740135 [FT] and No. 25800169 [TH]), and Inoue Foundation for Science.
This work was also supported by World Premier International Center Initiative
(WPI Program), MEXT, Japan [FT].

\appendix

\section{The tuning for the inflation in the string-inspired model}
\label{Sectuning}

In this section of the Appendix, we will discuss the tuning of
\bea
\hat{B} \approx \bigg(\frac{n_2}{n_1}\bigg)^2 \hat{A}
\eea
found in Sec.\ref{SECLET}.
Here, we define the phase of $B$:
\bea
\theta \equiv {\rm arg}[1/B].
\eea
It should be noted that the phase $\theta$ will be given by the VEVs of the dilaton and the complex structure moduli
stabilized by closed string fluxes. 
In terms of $A$ and $B$ this relation becomes
\bea
B \approx
A \bigg(\frac{n_2}{n_1}\bigg)^2 \bigg(\frac{W_0}{\log[1/W_0]}\bigg)^{-\frac{M}{2}\big(\frac{1}{n_2}-\frac{1}{n_1}\big)}
.
\eea
Here we substituted the solutions of moduli VEVs in Eq.(\ref{SolofT}).
For instance, one finds $B \sim 15 A$ for
$W_0 =10^{-4},~n_1 =7,~n_2=6$ and $M=22$. In this case, 
complex structure moduli can play a role in 1-loop threshold corrections from the heavy modes,
which depend on the gauge group.
On the other hand, for $W_0 =10^{-4},~n_1 =6,~n_2=3$ and $M=22$, one finds $B \sim 3.2\times 10^8 A$.
In the latter case, the heavy moduli such as the
dilaton and complex structure may play an important role in the relevant gauge couplings, e.g.,
\bea
\frac{4\pi}{g_1^2} \sim T_2,~~~\frac{4\pi}{g_2^2} \sim T_2 - \Delta f, 
\eea
where $\Delta f$ contains heavy moduli \cite{Blumenhagen:2006ci} and $ 1- \Delta f/T_2 \sim 1/2$.
Then one finds that $ B \sim e^{\frac{2\pi}{3}\Delta f} \sim e^{\frac{2\pi }{6}T_2} \gg 1$ \cite{Abe:2005rx}, 
using the fact that the size of one gaugino condensation is similar to the other,, 
$A e^{-2\pi T_2/6 } \sim B' e^{-2\pi (T_2-\Delta f)/3 }$, where $B' = B e^{-2\pi \Delta f/3 } ={\cal O}(1)$. 
Note that even if $B$ is much larger than unity, the heavy moduli/saxion stabilization does not change
as long as
\bea
\bigg(\frac{m_{\phi}}{m_{\sigma}}\bigg)^2 
\simeq \frac{|\hat{B}|}{|W_0|f^2} \lesssim 10^{-2} .
\eea 
A relation $M>2n_2$ is important to satisfy the above condition.

\section{Higher order terms in Choi-Jeong models}
\label{higherphi}

We write down higher order terms in $\Phi$:
\bea
K_L & \approx & 
\frac{f^2}{2} (\Phi +\Phi^{\dag})^2 
+ f^2 \frac{k_4}{4!} (\Phi +\Phi^{\dag})^4 + f^2 \frac{k_6}{6!} (\Phi +\Phi^{\dag})^6
+ \cdots , \\
W_L & \approx & W_0 + C e^{-\frac{\pi}{n_1}(\Phi + \langle T \rangle)}- D e^{-\frac{\pi}{n_2}(\Phi + \langle T \rangle)} 
 ,\\
V_{{\rm up}L} &\approx & 3m_{3/2}^2 + f^2 m_{3/2}^2 (\Phi +\Phi^{\dag})^2
+ 
f^2 m_{3/2}^2 \zeta (\Phi +\Phi^{\dag})^4
+ \cdots .
\eea
Here, we defined 
\bea
\nonumber
&& f^2 \equiv \frac{3}{2 \sqrt{2} \sqrt{t} {\cal V}},~~
k_4 \equiv \frac{9 \sqrt{2} t^{3/2}+6 {\cal V}}{8 t^2 {\cal V}}\sim f^2  <1 , ~~
k_6 \equiv \frac{15 \left(18 t^3+9 \sqrt{2} t^{3/2} {\cal V}+14 {\cal V}^2\right)}{32 t^4 {\cal V}^2} \sim f^4 <1, \\
&& m_{3/2} = e^{K/2}W = \frac{W_0}{{\cal V}},~~~
\zeta = \frac{ \left(7 t^{3/2}+\sqrt{2} {\cal V}\right) }{16 \sqrt{2} t^2 {\cal V}} \sim f^2 <1,
\eea
where $t = T+T^{\dag} $, ${\cal V} \equiv t_0^{3/2} - t^{3/2}  /\sqrt{2}$ and
$\hat{\epsilon} \approx 3{W_0}^2/{\cal V}^{2/3}$ are used, and it is noted that $t$ and ${\cal V}$ are given by the VEVs.
The higher order term of $\Phi$ in the K\"ahler potential will become irrelevant for $f \lesssim 1$, when 
the uplifting potential is added.

\end{document}